\documentclass[prd,twocolumn,superscriptaddress,showpacs]{revtex4}
\usepackage{latexsym}
\usepackage{amsmath}
\usepackage{amssymb}
\usepackage[dvips]{graphicx}
\newcommand{\calC}{\mathcal{C}}

\newcommand{\rd}{\mathrm{d}}
\begin{document}
%
%
%
\preprint{LAUR LA-UR-02-5413}
\title{Quantum dynamics of  phase transitions in
   broken symmetry $\lambda \, \phi^4$ field theory}
\author{Fred Cooper}
\email{fcooper@lanl.gov} \affiliation{Theoretical Division,
   Los Alamos National Laboratory, Los Alamos, NM 87545}
\author{John F. Dawson}
\email{john.dawson@unh.edu}
\affiliation{Department of Physics,
   University of New Hampshire, Durham, NH 03824}
\author{Bogdan Mihaila}
\email{bogdan.mihaila@unh.edu}
\affiliation{Physics Division,
   Argonne National Laboratory, Argonne, IL 60439}

\date{\today}
\begin{abstract}
We perform a detailed numerical investigation of the dynamics of a
single component  broken symmetry $\lambda \, \phi^4$ field theory
in 1+1 dimensions using a Schwinger-Dyson equation truncation
scheme based on ignoring vertex corrections.  In an earlier paper,
we called this the bare vertex approximation (BVA).  We assume the
initial state is described by a Gaussian density matrix peaked
around some non-zero value of $\langle \phi(0) \rangle$, and
characterized by a single particle Bose-Einstein distribution
function at a given temperature.  We compute the evolution of the
system using three different approximations: Hartree, BVA and a
related 2PI-1/N expansion, as a function of coupling strength and
initial temperature.  In the Hartree approximation, the static
phase diagram shows that there is a \emph{first} order phase
transition for this system.  As we change the initial starting
temperature of the system, we find that the BVA relaxes to a new
final temperature and exhibits a \emph{second} order phase
transition. We find that the average fields thermalize for
arbitrary initial conditions in the BVA, unlike the behavior
exhibited by the Hartree approximation, and we illustrate how
$\langle \phi(t) \rangle$ and $\langle \chi(t) \rangle$ depend on
the initial temperature and on the coupling constant.  We find
that the 2PI-1/N expansion gives dramatically different results
for $\langle \phi(t) \rangle$.
\end{abstract}
\maketitle
%
\newpage
%
%
\section{Introduction}
\label{s:introduction}

Recently there has been much effort in finding approximation
schemes to study the dynamics of phase transitions that go beyond
a leading order in large-N mean field theory approach.  This is an
important endeavor if one wants a first principles understanding
of the dynamics of quantum phase transitions.  In a previous set
of papers, we studied in quantum mechanics \cite{r:MCD01}, as well
as 1+1 dimensional field theory \cite{r:BCDM01}, the validity of a
$1/N$ motivated resummation scheme, which we called the bare
vertex approximation (BVA).

The long-term goal of this work is finding approximation schemes
which are accurate at the small values of N relevant to the case
of realistic quantum field theories: N=4 for the linear sigma
model, and N=2 for the Walecka model. In order to maximize the
possible differences between approximation schemes we choose to
study the O(N) model for $N=1$. Based on our previous studies of
the quantum mechanical version of this model~\cite{r:MCD01} we
expect that by increasing~N the differences will diminish.  This
is due to the fact that the Schwinger-Dyson formalism is related,
but not identical, with approximations based on the large-N
expansion.

This paper presents the first quantum-mechanical dynamical calculation
in 1+1 dimensions, using the BVA, of the explicitly broken symmetry
case, where the order parameter at $t=0$ is non-zero.  In 1+1
dimensions, it is known that there is no phase transition in this
model except at zero
temperature~\cite{r:Griffiths}. 
On the other hand, in two dimensional systems having
Berezinski-Kosterlitz-Thouless type
transitions~\cite{r:Berezinski70,r:KosterlitzThouless73,r:JKKN77},
the large-N expansion can give qualitatively good understanding of
the correlation functions, even when it gives the wrong phase
transition behavior~\cite{r:Witten78}. As the dimensions increase,
the mean field critical behavior becomes exact in four dimensions
and thus we expect that the approximation presented here should
improve as we increase the dimensionality. Thus we should think of
the model used here as a ``toy'' model for demonstrating some of
the features expected to be true in 3+1 dimensions such as the
restoration of symmetry breakdown at high temperatures and
thermalization of correlation functions. The model we are
ultimately interested in is the linear sigma model in 3+1
dimensions for the case of broken symmetry at finite temperature.
This model we studied earlier using a large-N approximation
\cite{r:KDC96,r:Lampertthesis,r:KCMPKnp95}.  Since the BVA
contains scattering contributions, it cures many of the problems
associated with mean field methods.  Here we are able to follow
the evolution of the system through a phase transition and study
the thermalization of the system.  We illustrate in this paper the
results of such calculations.

A parallel set of investigations by Berges et.~al.
\cite{r:BC01,r:AB01,r:AB02,r:B02,r:AABBS}, have looked at a related
approximation based on the two-particle irreducible expansion (which
they call 2PI-1/N).  These investigators have pointed out that when
there is broken symmetry, the BVA contains terms not included in the
$1/N$ resummation at next to leading order.  In this paper, we present
the first \emph{quantum} calculations which compare the BVA with the
2PI-1/N for the broken symmetry case.  Recently we were able to show
that for a \emph{classical} finite temperature $\lambda \phi^4$ field
theory in 1+1 dimensions, the BVA gave a better description of the
time evolution of $\langle \phi(t) \rangle$ than the 2PI-1/N expansion
and provided excellent agreement with exact Monte Carlo simulations
\cite{r:CDM02,r:BCDM01}.

In this paper we look at quantum evolutions in 1+1 dimensions,
starting with a Gaussian density matrix, and study how the evolution
of $\langle \phi(t) \rangle$ depends on the initial conditions and the
value of the coupling constant.  In the classical domain, the coupling
constant dependence can be scaled out, which is \emph{not} possible in
the quantum case we consider here.  Since we have not determined the
effective potential in the BVA approximation, we rely on the Hartree
approximation effective potential to guide our study.  The Hartree
potential however, indicates that the system should undergo a first
order phase transition.  In addition, in the Hartree approximation,
the fields never thermalize.  We find that the BVA cures these serious
problems.  In this paper, we show evolutions of the system as a
function of the initial ``temperature'' parameter and the coupling
constant, and, since in the BVA, the fields thermalize, we can follow
the system through what appears to be a second order phase transition.
We also compare our results with those of the 2PI-1/N expansion in the
quantum domain, and find that the two approximations diverge
dramatically for the behavior of $\langle \phi(t) \rangle$.

We derive the BVA equations for the general $N$-component $\lambda
\, [ \, \phi^2_i(x) \,]^2$ field theory in Sections
\ref{s:classactionandBVA} and \ref{s:updateGreen}.  We then
specialize to the case $N=1$, and in Section
\ref{s:hartreephasediagram}, we derive the Hartree approximation
phase diagram.  In Section \ref{s:initial}, we discuss the initial
conditions we choose for this problem.  Numerical results are
shown in Section \ref{s:numerical}, and conclusions discussed in
Section \ref{s:thermalization}.

%
%
\section{The classical action and time evolution in the BVA}
\label{s:classactionandBVA}

The classical action for $\lambda \phi^4$ with $N$ fields ($i=1
\ldots N$) is
\begin{align}
   S[\phi]
   =
   \int \rd^2 x \,
   \biggl \{ &
      \frac{1}{2} \,
         \Bigl [
            \partial_\mu \phi_i(x) \ \partial^\mu \phi_i(x)
            +
            \mu^2 \, \phi_i^2(x)
         \Bigr ]
\nonumber \\ &
      -
      \frac{\lambda}{8 N} \, [ \, \phi^2_i(x) \, ]^2
      -
      \frac{N \mu^4 }{2 \lambda} \,
   \biggr \} \>.
   \label{e:Lagi}
\end{align}
For the purposes of our resummation scheme which is motivated by
$1/N$ considerations it is useful to consider the alternative
action
\begin{align}
   S_{\text{cl}}[\phi_i, \chi]
   =
   \int \rd^2 x \,
   \biggl \{ &
   \frac{1}{2} \,
      \Bigl [
         \partial_\mu \phi_i(x) \ \partial^\mu \phi_i(x)
         - \chi(x) \, \phi_i^2(x)
      \Bigr ]
\nonumber \\ &
   +
   \frac{N}{\lambda} \,
      \Bigl [ \,
         \frac{\chi^2(x)}{2}
         +
         \mu^2 \, \chi(x) \,
      \Bigr ] \,
   \biggr \}  \>,
   \label{e:Lagii}
\end{align}
which leads to the Heisenberg equations of motion
\begin{equation}
   \left  [ \,
      \Box
      +
      \chi(x) \,
   \right ] \, \phi_i(x)
   = 0  \>,
   \label{e:eompclassical}
\end{equation}
and  the constraint (``gap'') equation for $\chi(x)$
\begin{equation}
   \chi(x)
   =
   - \mu^2
   +
   \frac{\lambda}{2 N} \, \phi_i(x)\phi_i(x) \>.
   \label{e:chiclassical}
\end{equation}
Throughout this paper, we use the Einstein summation convention for
repeated indices.

The BVA truncation scheme of the Schwinger-Dyson equations is most
easily obtained from the 2PI effective action \cite{r:CJT, r:LW,
r:Baym62}.  Other approaches leading to these equations are found in
\cite{r:MCD01,r:AABBS}.  Using the extended fields notation, $
\phi_\alpha(x) = [ \chi(x), \phi_1(x), \phi_2(x), \ldots , \phi_N(x)]
\>, $ the effective action for the evolution can be written as:
\begin{align}
   \Gamma[\phi_\alpha,G]
   \ = \ &
   S_{\text{cl}}[\phi_\alpha]
   +
   \frac{i}{2} {\rm Tr} \ln [ \, G^{-1} \, ]
   \notag \\ &
   +
   \frac{i}{2} {\rm Tr} [ \, G_0^{-1} \, G \, ]
   +
   \Gamma_2[G] \>,
   \label{e:BVAaction}
\end{align}
where $ \Gamma_2[G]$ is the generating functional of the 2-PI
graphs, and the classical action in Minkowski space is
\begin{align}
   S_{\text{cl}}[\phi_\alpha]
   =
   \int d^2x \,
   \biggl \{ &
   - \frac{1}{2}
   \phi_i(x) \,
   [ \, \Box + \chi(x) \, ] \, \phi_i(x)
   \notag \\ &
   +
   \frac{\chi^2(x)}{2 g}
   +
   \frac{\mu^2}{g} \chi(x) \,
   \biggr \} \>.
\end{align}
Here and in what follows we let $g=\lambda/N$.

The integrals and delta functions $\delta_{\calC}(x,x')$ are defined
on the closed time path (CTP) contour, which incorporates the initial
value boundary condition
\cite{r:Schwinger,r:Keldish,r:MahanI,r:MahanII}. The approximations we
are studying include only the two-loop contributions to $\Gamma_{2}$.

\begin{figure}[h!]
   \centering
   \includegraphics[width=2.in]{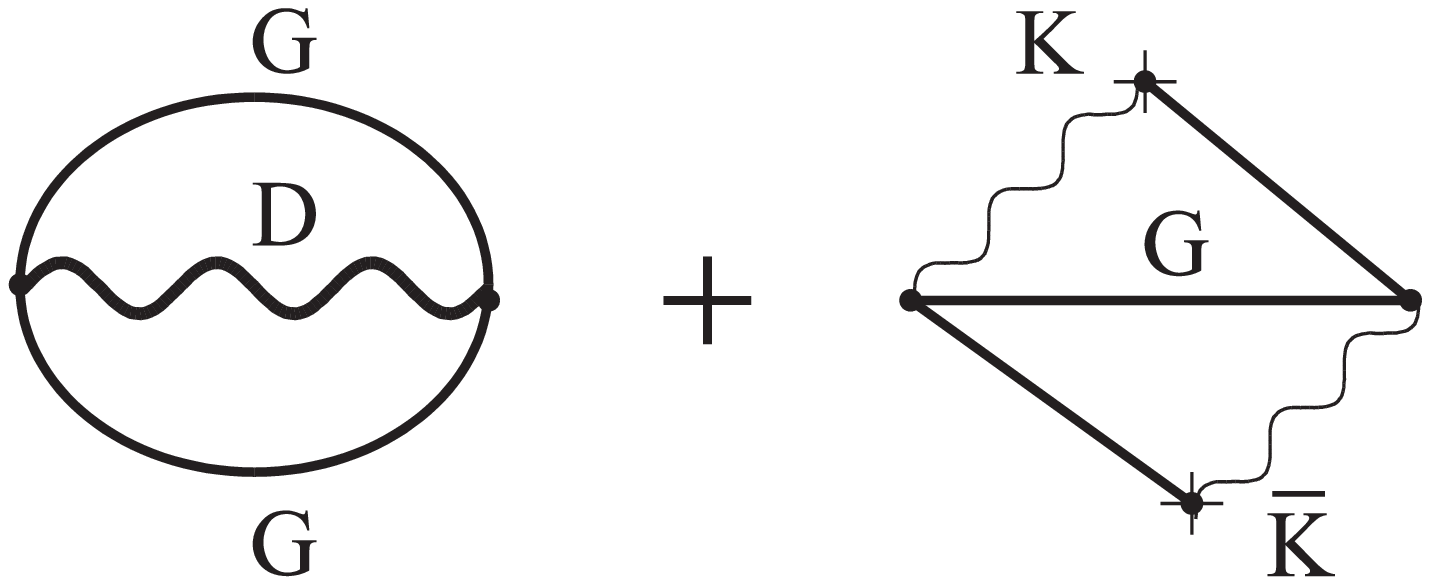}
   \caption{Graphs included in the 2PI effective action $\Gamma_2$.}
   \label{f:fig1}
\end{figure}

The Green function $G^{-1}_{0\,\alpha \beta}[\phi](x,x')$ is defined as follows:
\begin{eqnarray}
   G_{0\,\alpha \beta}^{-1}[\phi](x,x')
   & = &
   - \frac{\delta^2 S_{\text{cl}} }
          { \delta \phi_{\alpha}(x) \, \delta \phi_{\beta}(x') }
   \notag \\
   & = &
   \begin{pmatrix}
      D^{-1}_0(x,x')   & \bar{K}_{0\,j}^{-1}(x,x') \\
      K_{0\,i}^{-1}(x,x') & G_{0\,i j}^{-1}(x,x')
   \end{pmatrix}  \>,
   \label{e:ginvdef}
\end{eqnarray}
where
\begin{gather*}
   D_0^{-1}(x,x')
   =
   - \, g \, \delta_{\calC}(x,x') \>,
   \\
   G_{0\,ij}^{-1}[\chi](x,x')
   =
   [ \, \Box + \chi(x) \, ] \, \delta_{ij} \delta_{\calC}(x,x') \>,
   \\
   K_{0\,i}^{-1}[\phi](x,x')
   =
   \bar{K}_{0\,i}^{-1}[\phi](x,x')
   =
   \phi_i(x) \, \delta_{\calC}(x,x') \>.
\end{gather*}
The exact Green function $G_{\alpha \beta}[j](x,x')$ is defined
by:
\begin{equation}
   G_{\alpha \beta}[j](x,x')
   =
   \begin{pmatrix}
      D(x,x')     & K_j(x,x') \\
      \bar{K}_i(x,x') & G_{i j}(x,x')
   \end{pmatrix}  \>,
   \label{e:GGdef}
   \notag
\end{equation}
The exact equations following from the effective action
Eq.~\eqref{e:BVAaction}, are:
\begin{gather}
   [ \, \Box + \chi(x) \, ] \, \phi_i(x)
   +
   K_i(x,x) / i
   = 0 \>,
   \label{e:GammachieqBVA} \\
   \chi(x)
   = - \mu^2+
   \frac{g}{2}
      \sum_i
      \bigl [ \,
         \phi_i^2(x)
         +
         G_{ii}(x,x)/i \,
      \bigr ] \>,
   \notag
\end{gather}
and
\begin{equation}
   G_{\alpha \beta}^{-1}(x,x')
   =
   G_{0\, \alpha \beta}^{-1}(x,x')
   +
   \Sigma_{\alpha \beta}(x,x')  \>,
   \label{e:GGinvGinvSigma}
\end{equation}
where
\begin{eqnarray}
   \Sigma_{\alpha \beta}(x,x')
   & = &
   \frac{2}{i} \,
   \frac{\delta \Gamma_2[G]}{\delta G_{\alpha\beta}(x,x')}
   \notag \\
   & = &
   \begin{pmatrix}
      \Pi(x,x')          & \Omega_j(x,x') \\
      \bar\Omega_i(x,x') & \Sigma_{ij}(x,x')
   \end{pmatrix}
   \>.
   \label{e:Sigmasdefs}
\end{eqnarray}
In the BVA, we keep in $\Gamma_2[G]$ only the graphs shown in
Fig.~\ref{f:fig1}, which is explicitly
\begin{align}
   \label{e:Gamma2}
   \Gamma_2[G]
   =
   - \frac{1}{4}
   \iint \mathrm{d}x \mathrm{d}y
   \Bigl [ &
      G_{ij}(x,y) G_{ji}(y,x) D(x,y)
   \\ \notag &
      +
      2 \bar{K}_i(x,y) K_j(x,y) G_{ij}(x,y)
   \Bigr ]
   \>.
\end{align}
The self-energy, given in Eq.~\eqref{e:Sigmasdefs}, then reduces
to:
\begin{align}
   \Pi(x,x')
   &=
      \frac{i}{2} \,
      G_{mn}(x,x') \, G_{mn}(x,x')
   \>,
   \label{e:SigmasBVA} \\
   \Omega_i(x,x')
   &=
      i \,
      \bar{K}_m(x,x') \, G_{m i}(x,x')
   \>,
   \notag \\
   \bar\Omega_i(x,x')
   &=
      i \,
      \bar{K}_m(x,x') \, G_{m i}(x,x')
   \>,
   \notag \\
   \Sigma_{ij}(x,x')
   &=
      i \,
      [ \,
         G_{i j}(x,x')\, D(x,x')
         +
         \bar{K}_{i }(x,x')\, K_{j}(x,x') \,
      ]
   \,.
   \notag
\end{align}

As discussed in detail in Ref.~\cite{r:AABBS}, the second graph in
Fig.~\ref{f:fig1} is proportional to $1/N^2$ and is ignored in the
2PI-1/N expansion. Our recent simulations in the classical domain
showed that the BVA gave a more accurate determination of $\langle
\phi(t) \rangle$, and we will concentrate in this paper on the BVA
except to point out with explicit results that in the quantum
domain the differences between the BVA and the 2PI-1/N expansion
grow with increasing coupling constant $g$ (for the case $N=1$
studied here) and that unlike the BVA, the 2PI-1/N expansion does
not track the average of the Hartree result.

%
%
\section{Update equations for the Green functions}
\label{s:updateGreen}

We notice from the definitions of the matrices representing
$G_{\alpha\beta}(x,x')$ and $G_{\alpha\beta}^{-1}(x,x')$, that the
matrix elements are not inverses of one another, but instead
satisfy schematically:
\begin{align}
   D^{-1} D
    + \bar{K}_k^{-1} \bar{K}_k
   &= \delta_{\calC} \>,
   \label{e:GinvG}
   \\ \notag
   K_i^{-1} D + G_{ik}^{-1} \bar{K}_k
   &= 0
   \\ \notag
   D^{-1} K_j + \bar{K}_k^{-1} G_{kj}
   &= 0 \>,
   \\ \notag
   K_i^{-1} K_j
    + G_{ik}^{-1} G_{kj}
   &=
   \delta_{ij} \delta_{\calC} \>.
\end{align}
Inverting Eq.~\eqref{e:GGinvGinvSigma}, we find:
\begin{widetext}
\begin{eqnarray}
   D(x,x')
   & = &  - g \, \delta_{\calC}(x,x')
          + g \, \int_{\calC} \mathrm{d}x_1 \ \Pi'(x,x_1) \
   D(x_1,x')
   \>,
   \\
   G_{ij}(x,x')
   & = &
   G_{0\,ij}(x,x') \delta_{ij}
   - \int_{\calC} \mathrm{d}x_1 \int_{\calC} \mathrm{d}x_2 \
   G_{0\,ik}(x,x_1) \, \Sigma_{kl}'(x_1,x_2) \
   G_{lj}(x_2,x')
   \>,
   \\
   K_i(x,x')
   & = & - \int_{\calC} \mathrm{d}x_1 \int_{\calC} \mathrm{d}x_2 \
           \Bigl [ D_0^{-1} + \Pi \Bigr ]^{-1}\!\! (x,x_1) \,
           \Bigl [ \bar K_{0\,k}^{-1} + \Omega_k \Bigr ] (x_1,x_2) \,
           G_{ki}(x_2,x')
   \>,
\end{eqnarray}
with
\begin{eqnarray}
   \Pi'(x,x')
   & = &
   \Pi(x,x')
   - \int_{\calC} \mathrm{d}x_1 \int_{\calC} \mathrm{d}x_2 \,
      \Bigl [ \bar{K}_{0\,k}^{-1} + \Omega_k \Bigr ](x,x_1) \,
      \Bigl [ G_{0\,kl}^{-1} + \Sigma_{kl} \Bigr ]^{-1}\!\!(x_1,x_2) \,
      \Bigl [ K_{0\,l}^{-1} + \bar{\Omega}_l \Bigr ](x_2,x')
   \>,
   \label{e:Sigmap}
   \\
   \Sigma_{ik}'(x,x')
   & = &
   \Sigma_{ik}(x,x')
    - \int_{\calC} \mathrm{d}x_1 \int_{\calC} \mathrm{d}x_2
      \Bigl [ K_{0\,i}^{-1} + \bar{\Omega}_i \Bigr ](x,x_1) \
      \Bigl [ D_0^{-1} + \Pi \Bigr ]^{-1}\!\!(x_1,x_2) \
      \Bigl [ \bar{K}_{0\,k}^{-1} + \Omega_k \Bigr ](x_2,x')
   \>.
\end{eqnarray}
\end{widetext}
These update equations must be solved in conjunction with the
one-point functions, Eqs.~\eqref{e:GammachieqBVA}.

For a practical implementation of the above approach we need to
solve for $D_2(x,x')$ and $G_{2, ij} (x,x')$, the inverses of
$\Bigl [ D_0^{-1} + \Pi \Bigr ] (x,x')$ and $[ G_{ij}^{-1} +
\Sigma_{ij} \Bigr ] (x,x')$, respectively. We have
\begin{align}
   D_2(x,x')
   = & - g \delta_{\calC}(x,x')
   \\ \notag &
     + g \int_{\calC} \mathrm{d}x_1 \Pi(x,x_1) D_2(x_1,x')
   \>,
\end{align}
\begin{align}
   & G_{2, ij}(x,x')
   =
   G_{0\, ij}(x,x')
   \\ \notag & \qquad
   - \int_{\calC} \mathrm{d}x_1 \int_{\calC} \mathrm{d}x_2
   G_{0\, ik}(x,x_1)
   \Sigma_{kl}(x_1,x_2) G_{2, lj}(x_2,x')
   \>.
\end{align}
We also perform the following substitutions
\begin{eqnarray}
   D(x,x') & = & - g \, \delta_{\calC}(x,x') + \bar{D}(x,x')
   \>,
   \\ \nonumber
   D_2(x,x') & = & - g \, \delta_{\calC}(x,x') + \bar D_2(x,x')
   \>,
   \\ \nonumber
   K_i(x,x') & = & g \, \phi_k(x) G_{ki}(x,x') + \bar K_i(x,x')
   \>.
\end{eqnarray}
Thus we obtain the equations of motion
\begin{gather}
   \Bigl \{
      [ \, \Box + \chi(x) \, ] \delta_{ik}
      + g \, G_{ki}(x,x) / i
   \Bigl \} \phi_k(x)
   +
   \bar K_i(x,x) / i
   = 0 \>,
   \notag \\
   \chi(x)
   =
   - \mu^2
   +
   \frac{g}{2} \, \phi_i^2(x)
   +
   \frac{g}{2} \, [ \, G_{ii}(x,x)/i \, ]
   \>.
\end{gather}
and the update equations for the Green functions
\begin{widetext}
\begin{eqnarray}
   \bar D(x,x')
   & = &  - g^2 \, \Pi'(x,x_1)
          + g \, \int_{\calC} \mathrm{d}x_1 \ \Pi'(x,x_1) \bar D(x_1,x')
   \>,
\end{eqnarray}
\begin{eqnarray}
   \bar D_2(x,x')
   & = &  - g^2 \, \Pi(x,x_1)
          + g \, \int_{\calC} \mathrm{d}x_1 \ \Pi(x,x_1) \bar D_2(x_1,x')
   \>,
\end{eqnarray}
\begin{eqnarray}
   G_{ij}(x,x')
   & = &
   G_{0\, ij}(x,x')
   - g \int_{\calC} \mathrm{d}x_1 \
   G_{0\, ik}(x,x_1)
     \Bigl [ \phi_k(x_1) \phi_l(x_1) + G_{k l}(x_1,x_1)/i
     \Bigr ] G_{lj}(x_1,x')
   \\ \nonumber &&
   - \int_{\calC} \mathrm{d}x_1 \int_{\calC} \mathrm{d}x_2 \
   G_{0\, ik}(x,x_1) \bar \Sigma_{kl}'(x_1,x_2) G_{lj}(x_2,x')
   \>,
   \\
   G_{2\, ij}(x,x')
   & = &
   G_{0\, ij}(x,x')
   - g \int_{\calC} \mathrm{d}x_1 \
   G_{0\, ik}(x,x_1)
     [ \, G_{2\, k l}(x_1,x_1)/i \, ] G_{2\, lj}(x_1,x')
   \\ \nonumber &&
   - \int_{\calC} \mathrm{d}x_1 \int_{\calC} \mathrm{d}x_2 \
   G_{0\, ik}(x,x_1) \bar \Sigma_{kl}(x_1,x_2) G_{lj}(x_2,x')
   \>,
\end{eqnarray}
and
\begin{eqnarray}
   \bar K_i(x,x')
   & = &
   - \int_{\calC} \mathrm{d}x_1 \ \bar D_2(x,x_1) \phi_k(x_1) G_{ki}(x_1,x')
   \\ \nonumber &&
   +
   g \int_{\calC} \mathrm{d}x_1 \ \Omega_k(x,x_1) G_{ki}(x_1,x')
   - \int_{\calC} \mathrm{d}x_1 \int_{\calC} \mathrm{d}x_2 \
           \bar D_2(x,x_1) \Omega_k(x_1,x_2) G_{ki}(x_2,x')
   \>,
\end{eqnarray}
with
\begin{eqnarray}
   \Pi'(x,x')
   & = &
   \Pi(x,x')
   - \phi_k(x) G_{2,kl}(x,x') \phi_l(x')
   - \int_{\calC} \mathrm{d}x_1 \int_{\calC} \mathrm{d}x_2 \
     \Omega_k(x,x_1) G_{2,kl}(x_1,x_2) \bar \Omega_l(x_2,x')
   \\ \nonumber &&
   - \int_{\calC} \mathrm{d}x_1 \ \phi_k(x) G_{2,kl}(x,x_1) \bar \Omega_l(x_1,x')
   - \int_{\calC} \mathrm{d}x_1 \ \Omega_k(x,x_1) G_{2,kl}(x_1,x') \phi_l(x')
   \>,
   \\
   \bar \Sigma_{ik}(x,x')
   & = &
   i \,
      [
        G_{i k}(x,x') \, \bar D(x,x')
         +
         \bar K_i(x,x') \, K_k(x,x')
      ]
   \>,
   \\
   \bar \Sigma_{ik}'(x,x')
   & = &
   \bar \Sigma_{ik}(x,x')
   - \phi_i(x) \bar D_2(x,x') \phi_k(x')
   - \int_{\calC} \mathrm{d}x_1 \int_{\calC} \mathrm{d}x_2 \
     \bar \Omega_i(x,x_1) \bar D_2(x_1,x_2) \Omega_k(x_2,x')
   \\ \nonumber &&
   - \int_{\calC} \mathrm{d}x_1 \ \phi_i(x) \bar D_2(x,x_1) \Omega_k(x_1,x')
   - \int_{\calC} \mathrm{d}x_1 \ \bar \Omega_i(x,x_1) \bar D_2(x_1,x')
     \phi_k(x')
   \\ \nonumber &&
   + g \,
   \Bigl [ \phi_i(x) \Omega_k(x,x') + \bar \Omega_i(x,x') \phi_k(x')
   \Bigr ]
   + g
   \int_{\calC} \mathrm{d}x_1 \ \bar \Omega_i(x,x_1) \, \Omega_k(x_1,x')
   \>.
\end{eqnarray}
For computational purposes, it is suitable to make one more
transformation of the $\phi_i$, $G_{ij}$ and $G_{2\, ij}$
equations. We write the equivalent integro-differential equation
for $G_{ij}$ as:
\begin{multline}
   \Bigl \{
      [ \, \Box + \chi(x) \, ] \delta_{ik}
      +
      g \, [
         \phi_i(x) \phi_k(x) + G_{i k}(x,x)/i
           ]
   \Bigr \} G_{k j}(x,x')
   =
   \delta_{\calC}(x,x') \, \delta_{ij}
   - \int_{\calC} \mathrm{d}x_1 \,
   \bar \Sigma_{i k}(x,x_2) \, G_{k j}(x_2,x')
   \>.
\end{multline}
\end{widetext}
We specialize now to the case $N=1$. It is convenient then to
introduce the following equations
\begin{equation}
   [ \, \Box + \chi_1(x) \, ] \, \phi(x)
   +
   \bar K(x,x) / i
   = 0 \>,
\end{equation}
\begin{equation}
   \chi_1(x)
   =
   - \mu^2
   +
   \frac{g}{2} \, \phi^2(x)
   +
   \frac{3 g}{2} \, G(x,x)/i
   \>,
\end{equation}
together with redefinitions for $G_{0}(x,x')$ to work with the
equations for $G(x,x')$ and $G_{2}(x,x')$, respectively. We have
\begin{gather}
   [ \, \Box + \chi_2(x) \, ] \,
   \bar G_{0}(x,x')
   =
   \delta_{\calC}(x,x')
   \\
   [ \, \Box + \chi_1(x) \, ] \,
   \bar {\bar G}_{0}(x,x')
   =
   \delta_{\calC}(x,x')
   \>,
\end{gather}
with
\begin{gather}
   \chi_2(x) =
   - \mu^2
   + \frac{3 g}{2} \,
     \bigl [ \, \phi^2(x) +  G(x,x)/i \, \bigr ]
   \>,
\end{gather}
Finally, the modified equations are given by:
\begin{align}
   &G(x,x')
   =
   \bar G_{0}(x,x')
   \\ \notag & \quad
   - \int_{\calC} \mathrm{d}x_1 \int_{\calC} \mathrm{d}x_2
   \bar G_{0}(x,x_1)
   \bar \Sigma'(x_1,x_2)
   G(x_2,x')
   \>,
\end{align}
\begin{align}
   &G_{2}(x,x')
   =
   \bar {\bar G}_{0}(x,x')
   \\ \notag & \quad
   - \int_{\calC} \mathrm{d}x_1 \int_{\calC} \mathrm{d}x_2
   \bar {\bar G}_{0}(x,x_1)
   \bar \Sigma(x_1,x_2)
   G_{2}(x_2,x')
   \>.
\end{align}

%
%
\section{Hartree Phase Diagram}
\label{s:hartreephasediagram}

It would be useful to have available the effective potential for the
BVA approximation from Eq.~\eqref{e:BVAaction} to use as a guide for
starting out the BVA solutions.  However solving the self-consistent
equations of the BVA and constructing the thermal effective potential
is a formidable task, and has only been recently considered for the
simpler loop approximation to $\lambda \phi^4$ for the single field
($N=1$) case (see Refs.~\cite{r:vanHees1, r:vanHees2}).  Therefore in
this paper, we consider the effective potential for the simpler
Hartree approximation for a single field.

The effective action in the Hartree approximation can be written
in the form
\begin{align}
   S[\phi,\chi]
   = \ &
   \int \rd^2 x \,
   \biggl \{
      - \frac{1}{2} \,
      \phi \, [ \, \Box + \chi \, ] \, \phi
      +
      \frac{\lambda}{4} \phi^4
   \\ \notag & \quad
      +
      \frac{1}{3\lambda}
      \left (
         \frac{\chi^2}{2} + \mu^2 \chi
      \right )
   +
   \frac{i}{2} \mathrm{Tr} [ \ln ( \, \Box + \chi \, ) ]
   \biggr \}
    \>.
\end{align}
This action gives the Hartree equations of motion:
\begin{gather}
   [ \, \Box + \chi(x) - \lambda \, \phi^2(x) \, ] \, \phi(x)
   =
   0 \>,
   \label{e:Hartreefe} \\
   \chi(x)
   =
   - \mu^2
   +
   \frac{3 \lambda}{2} \,
      \phi^2(x)
      +
   \frac{3 \lambda}{2} \,
      \mathrm{Tr} \, [ \, G_0/i \, ]
   \>,
   \notag
\end{gather}
with $G_0^{-1} (x,x') = [ \, \Box + \chi(x) \, ] \, \delta(x,x')$.

The effective potential for this action is given by:
\begin{align*}
   & V_{\text{H}}[\phi,\chi]
   = \
   V_{\text{cl}}[\phi,\chi]
   \\ & \qquad
   +
   \int_{0}^{+\infty} \frac{\rd k}{2\pi}
   \Bigl \{
      \omega_{k}
      +
      \frac{2}{\beta} \ln [ 1 - \exp ( -\beta \omega_{k} ) ]
   \Bigr \}  \>,
   \\
   &
   V_{\text{cl}}[\phi,\chi]
   =
   \frac{1}{2} \, \chi \, \phi^2
   -
   \frac{1}{3 \lambda} \,
   \Bigl ( \,
      \frac{\chi^2}{2} + \mu^2 \, \chi \,
   \Bigr )
   -
   \frac{\lambda}{4} \phi^4  \>,
\end{align*}
where $\omega_k = \sqrt{ k^2 + \chi}$.  We note that the requirement:
\begin{equation*}
   \frac{ \partial V_{\text{H}}[\phi] }{\partial \chi}
   =
   \frac{\phi^2}{2}
   -
   \frac{1}{3 \lambda} ( \chi + \mu^2 )
   +
   \int_{0}^{+\infty} \frac{\rd k}{2\pi}
   \frac{ 2 n_k + 1}{ 2 \omega_k}
   = 0  \>,
\end{equation*}
leads to the gap equation:
\begin{equation*}
   \chi
   =
   - \mu^2
   +
   \frac{3\lambda}{2} \,
      \phi^2
      +
   \frac{3\lambda}{2} \,
      \int_{0}^{+\infty} \frac{\rd k}{2\pi} \,
      \frac{ 2 n_k + 1}{\omega_k}
   \>.
\end{equation*}
where $n_k = 1/[ e^{\beta \omega_k} - 1]$.  The above equations are
infinite, so to renormalize them, we introduce a cutoff at $k = \pm
\Lambda$, and introduce a quantity $m^2 > 0$, defined by:
\begin{align}
   \label{e:mtomu}
   - m^2
   = \ &
   - \mu^2
   +
   \frac{3\lambda}{2} \,
   \int_{0}^{\Lambda} \frac{\rd k}{2\pi} \,
   \frac{1}{\sqrt{k^2 + m^2}}
   \\ \notag \approx \ &
   - \mu^2
   +
   \frac{3\lambda}{4\pi} \,
   \ln ( \, \Lambda / m \, )  \>.
\end{align}
Recall that $\mu^2 > 0$.  Subtracting \eqref{e:mtomu} from the gap
equation gives:
\begin{align}
\label{e:rengap}
   \chi
   = \ &
   - m^2
   +
   \frac{3\lambda}{2} \,
      \phi^2
   \\ \notag &
   +
   \frac{3\lambda}{2} \,
      \int_{0}^{\Lambda} \frac{\rd k}{2\pi} \,
   \biggl ( \,
      \frac{2 n_k + 1}{\sqrt{k^2 + \chi}}
      -
      \frac{1}{\sqrt{k^2 + m^2}} \,
   \biggr )
   \>,
   \\
   \approx \ &
   - m^2
   +
   \frac{3\lambda}{2} \,
   \biggl [ \,
      \phi^2
      +
      \frac{1}{4\pi} \, \ln ( m^2 / \chi )
      +
      \int_{0}^{\Lambda} \frac{\rd k}{\pi} \,
      \frac{n_k}{\omega_k} \,
   \biggr ]
   \>,
   \notag
\end{align}
which is now finite.  The Hartree potential, is renormalized at
$T=0$ by first renormalizing the partial derivative:
\begin{align*}
   \frac{\partial V_{\text{H}}[\phi,\chi]}{\partial \chi}
   = \ &
   \frac{\phi^2}{2}
   -
   \frac{1}{3\lambda} \, ( \, \chi + m^2 \, )
   \\ &
   +
   \int_{0}^{\Lambda} \frac{\rd k}{2\pi} \,
   \biggl ( \,
      \frac{1}{2 \sqrt{k^2 + \chi}}
      -
      \frac{1}{2 \sqrt{k^2 + m^2}} \,
   \biggr )  \>,
   \\
   \approx \ &
   \frac{\phi^2}{2}
   -
   \frac{1}{3\lambda} \, ( \, \chi + m^2 \, )
   +
   \frac{1}{8\pi} \, \ln ( m^2/\chi ) \>,
\end{align*}
where we have used \eqref{e:mtomu} to make the equation finite.
Partially integrating, we obtain the renormalized Hartree effective
potential:
\begin{align}
   V_{\text{H}}[\phi,\chi]
   = \ &
   \frac{1}{2} \, \chi \, \phi^2
   -
   \frac{\lambda}{4} \, \phi^4
   -
   \frac{1}{3\lambda} \,
   \Bigl ( \,
      \frac{\chi^2}{2} + m^2 \, \chi \,
   \Bigr )
   \\ \notag &
   +
   \frac{1}{8\pi} \,
   \Bigl [ \chi - \chi \, \ln ( m^2/\chi ) \Bigr ]
   \\ \notag &
   +
   \int_{0}^{\infty} \frac{\rd k}{\pi} \,
       \frac{1}{\beta} \ln [ 1 - \exp ( -\beta \omega_{k} ) \, ]
   \>,
   \label{e:VHeff}
\end{align}
where we have added back in the finite temperature-dependent part.
This equation is to be solved with $\chi$ satisfying the renormalized
gap equation \eqref{e:rengap}.  In practice, it is useful to solve
both of these equations parametrically as a function of $\chi$.

\begin{figure}[h!]
   \includegraphics[width=3.0in]{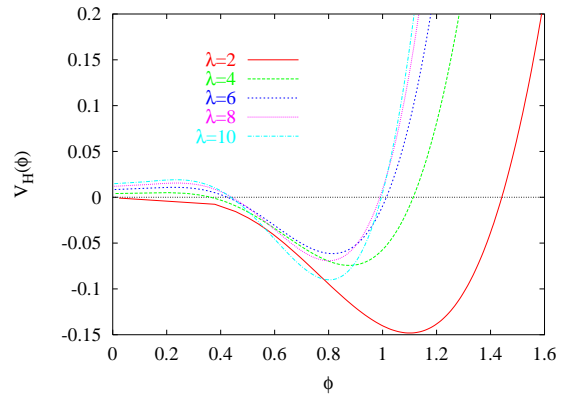}
   \caption{\label{fig:vpot_lambda}We set $T=1$, and plot
            the Hartree effective potential as a function of
            the coupling constant.}
\end{figure}

\begin{figure}[h!]
   \includegraphics[width=3.0in]{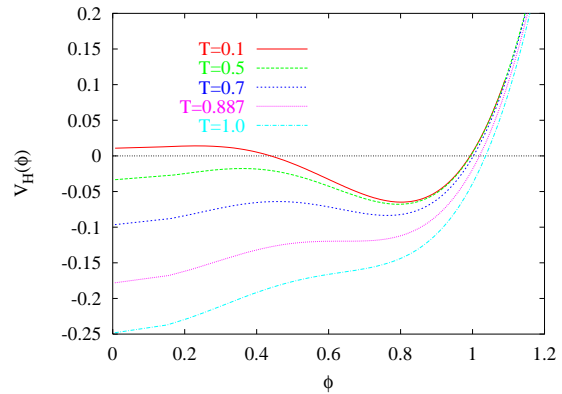}
   \caption{\label{fig:vpot_temp}We set $\lambda=7.3$, and plot
            the Hartree effective potential as a function of
            temperature.}
\end{figure}

The physical (renormalized) mass is given by the second derivative
of the effective potential, evaluated at the minimum.  The minimum
occurs at
\begin{equation*}
   \biggl [ \,
   \frac{ \rd V_{\text{H}}[\phi,\chi] }
        { \rd \phi }
   \biggr ]_{\phi = v}
   =
   v \, ( \chi - \lambda \, v^2 )
   =
   0  \>.
\end{equation*}
This implies that for the symmetry-breaking solution, $\chi = \lambda
v^2$.  Thus, from the gap equation \eqref{e:rengap}, the position of
the minimum and the mass parameter $m^2$ are related by:
\begin{equation}
   m^2
   =
   \frac{\lambda}{2} v^2
   +
   \frac{3\lambda}{2}
   \biggl [
      \frac{1}{4\pi} \ln \Bigl ( \frac{m^2}{\lambda v^2} \Bigr )
      +
      \int_{0}^{\Lambda} \frac{\rd k}{\pi}
         \frac{ n_{k}(\lambda v^2) }{\omega_k(\lambda v^2)}
   \biggr ]  \>.
   \label{e:mandv}
\end{equation}
The renormalized mass $m_{\text{R}}^2$ is defined by:
\begin{equation*}
   m_{\text{R}}^2
   =
   \biggl [ \,
   \frac{ \rd^2 V_{\text{H}}[\phi,\chi] }
        { \rd \phi^2 }
   \biggr ]_{\phi = v}
   =
   \biggl [ \,
      \chi
      -
      3 \lambda \phi^2
      +
      \phi \frac{\partial \chi}{\partial \phi}
   \biggr ]_{\phi = v}
   \>.
\end{equation*}
>From the gap equation \eqref{e:rengap}, we find:
\begin{equation*}
   \biggl [ \,
      \frac{\partial \chi}{\partial \phi} \,
   \biggr ]_{\phi = v}
   =
   \frac{3 \lambda v}
   { 1 + 3  [ \, 1 + f(\lambda v^2) \, ] / ( 8\pi v^2) } \>,
\end{equation*}
where $f(\lambda v^2)$ is the finite integral:
\begin{equation*}
   f(\chi)
   =
   \chi
   \int_{0}^{\Lambda} \rd k \,
      \frac{2 n_k(\chi)}{\omega_k^3(\chi)} \,
      \biggl [ \,
         1
         +
         \beta \, \omega_k(\chi) \, n_k(\chi) \,
            e^{\beta\omega_k(\chi)}
      \biggr ]  \>.
\end{equation*}
Thus the renormalized mass can be computed from:
\begin{equation*}
   m_{\text{R}}^2
   =
   \lambda \,
   \frac{ v^2 - 3 [ \, 1 + f(\lambda v^2) \, ] / ( 4\pi)}
   { 1 + 3  [ \, 1 + f(\lambda v^2) \, ] / ( 8\pi v^2) } \>,
\end{equation*}
The critical temperature $T_{\text{cr}}$ is defined by the
simultaneous solutions of:
\begin{equation*}
   v^2
   =
   \frac{3}{4\pi} \,
   \bigl [ \,
      1 + f(\lambda v^2) \,
   \bigr ] \>,
\end{equation*}
and Eq.~\eqref{e:mandv}. At $T=0$, we notice that unless $v^2 >
3/(4\pi)$, one cannot have a symmetry-breaking solution in this
approximation.  The effective potential as a function of
temperature~$T$ can be computed numerically.  In
Fig.~\ref{fig:vpot_lambda} we depict the effective potential
dependence on the coupling constant at a fixed temperature,
$T=0.1$. In Fig.~\ref{fig:vpot_temp} we fix the coupling constant at
$\lambda=7.3$, and depict the dependence of the effective potential on
the temperature. This particular value of $\lambda$ was used in our
study of the dynamics of disoriented chiral condensates (DCC) in $3+1$
dimensions in the leading order in large-N
approximation~\cite{r:KDC96,r:KCMPKnp95}. Here, the phase transition
occurs with $T_{\text{cr}} \approx 0.878$. We see that the phase
transition is first order with the vacuum value $v \approx 0.635$. The
value of $\chi$ at this vacuum is $\lambda v^2 \approx 2.94$.

%
%
\section{Initial Conditions}
\label{s:initial}

For the purpose of this study, we will assume that initially (at
$t=0$) the system is described by a Gaussian density matrix. Thus
initially the field equation and two-point equation is that of the
Hartree approximation.  The field equation obeys
\eqref{e:Hartreefe}, with $G_0(x,x')$ satisfying:
\begin{equation}
   [ \, \Box + \chi(t) \, ] \, G_0(x,x')
   =
   \delta_{\mathcal{C}}(x,x') \>.
\end{equation}
We can solve this Green function equation by introducing a set of
quantum fields $\phi_0(x)$, satisfying canonical commutation
relations $[ \, \phi_0(x), \dot{\phi}_0(x') \, ] = i \,
\delta(x-x') \>$, and obeying the homogeneous differential
equation
\begin{equation}
   [ \, \Box + \chi(t) \, ] \, \phi_0(x)
   =
   0 \>.
\end{equation}
In terms of these fields we have
\begin{align}
   G_0(x,x')
   &=
   i \, \langle \, \mathcal{T}_{\mathcal{C}} \{ \,
      \phi_0(x) \, \phi_0(x') \, \} \rangle \>,
   \notag \\
   &=
   G_{>}(x,x') \, \Theta_{\mathcal{C}}(t,t')
   +
   G_{<}(x,x') \, \Theta_{\mathcal{C}}(t',t) \>,
   \notag \\
   &=
   \int_{-\infty}^{+\infty} \frac{\rd k}{2\pi} \,
   \tilde{G}_{0}(k;t,t') \, e^{i k(x - x') }  \>.
\end{align}
We next expand these operators in Fourier mode functions
\begin{equation}
   \phi_0(x)
   =
   \int_{-\infty}^{+\infty} \frac{ \rd k}{2\pi} \,
   \Bigl [ \,
      a_{0\,k}^{\phantom\dagger} \, f_k^{\phantom\ast}(t) \, e^{i k x }
      +
      a_{0\,k}^{\dagger} \, f_k^{\ast}(t) \, e^{-i k x } \,
   \Bigr ]  \>,
\end{equation}
where the mode functions $f_k^{\phantom\ast}(t)$ satisfy:
\begin{equation}
   [ \, \partial_t^2 + \omega_k^2(t) \, ] \, f_k(t) = 0  \>,
   \qquad
   \omega_k(t) = \sqrt{ k^2 + \chi(t) }  \>,
\end{equation}
and the Wronskian condition: $ f_k^{\ast}(t) \,
\dot{f}_k^{\phantom\ast}(t) - \dot{f}_k^{\ast}(t) \,
f_k^{\phantom\ast}(t) = -i$. The operators
$a_{0\,k}^{\phantom\dagger}$ and $a_{0\,k}^{\dagger}$ satisfy the
usual commutation relations: $ [ \, a_{0\,k}^{\phantom\dagger},
a_{0\,k'}^{\dagger} \, ] = 2\pi \, \delta(k - k')$.  We will take
our initial density matrix such that:
\begin{align}
   \langle \,
      a_{0\,k}^{\dagger} \, a_{0\,k'}^{\phantom\dagger} \,
   \rangle
   &=
   n_k \, 2\pi \, \delta(k - k') \>,
   &
   \langle \,
      a_{0\,k}^{\phantom\dagger} \, a_{0\,k'}^{\phantom\dagger} \,
   \rangle
   &=
   0 \>,
   \notag \\
   \langle \,
      a_{0\,k}^{\phantom\dagger} \, a_{0\,k'}^{\dagger} \,
   \rangle
   &=
   ( n_k + 1 )\, 2\pi \, \delta(k - k') \>,
   &
   \langle \,
      a_{0\,k}^{\dagger} \, a_{0\,k'}^{\dagger} \,
   \rangle
   &=
   0 \>,
\end{align}
where $n_k = 1 / [ \, \exp( \beta_0 \omega_k(0) ) - 1 \, ]$.  Here $T_0 =
1/\beta_0$ is just a parameter for the initial Gaussian density
distribution, and is not the \emph{true} temperature of the
interacting system.  In fact, the system will not be in equilibrium at
$t=0$, but will arrive at a final temperature $T$ after it has come to
equilibrium.

Solutions for the mode functions $f_k^{\phantom\ast}(t)$ are of the
form:
\begin{equation}
   f_k^{\phantom\ast}(t)
   =
   \frac{e^{-i \int_0^t \Omega_k(t') \rd t' }}{\sqrt{2 \Omega_k(t)}} \>,
\end{equation}
where $\Omega_k(t)$ satisfies the non-linear differential equation,
\begin{equation}
   \frac{1}{2}
   \left  (
      \frac{ \ddot\Omega_k(t) }{ \Omega_k(t) }
   \right )
   -
   \frac{3}{4}
   \left  (
      \frac{ \dot\Omega_k(t) }{ \Omega_k(t) }
   \right )^{\!2}
   +
   \Omega^2_k(t)
   =
   \omega^2_k(t) \>.
   \label{e:Omega}
\end{equation}
The first order WKB solution for $f_k^{\phantom\ast}(t)$ is then given by
$\Omega_k(t) = \omega_k(t)$.  We take these solutions for our initial
conditions, so that at $t=0$,
\begin{alignat}{2}
   \Omega_k(0)
   &=
   \omega_k(0)
   &&=
   \sqrt{ k^2 + \chi(0) } \>,
   \notag \\
   \dot{\Omega}_k(0)
   &=
   \dot{\omega}_k(0)
   &&=
   \dot{\chi}(0) / 2 \omega_k(0)  \>.
   \label{e:initialomegas}
\end{alignat}
This means that
\begin{align}
   f_k(0)
   &=
   1 / \sqrt{ 2 \omega_k(0) } \>,
   \notag \\
   \dot{f}_k(0)
   &=
   - \left  [
      \frac{ \dot\omega_k(0) }{ 2 \omega_k(0) }
      +
      i \omega_k(0)
     \right ] \, f_k(0)  \>.
   \label{e:initialfq0}
\end{align}
We still need to find the value of $\chi_0(0)$.  This is given by
the Hartree self-consistent solutions of:
\begin{align}
   \chi(0)
   = &
   - \mu^2
   +
   \frac{3 \lambda}{2} \,
      \phi^2(0)
      +
   \frac{3 \lambda}{2} \,
      \int_{0}^{+\infty} \frac{\rd k}{2\pi} \,
     \frac{ 2 n_k + 1 }{ \sqrt{ k^2 + \chi(0) } }
   \>,
   \notag \\
   \label{e:chibar0}
   = &
   - m^2
   +
   \frac{3 \lambda}{2} \,
      \phi^2(0)
   \\ \notag &
   +
   \frac{3 \lambda}{2} \,
      \int_{0}^{\Lambda} \frac{\rd k}{2\pi} \,
      \left [ \,
         \frac{ 2 n_k + 1 }{\sqrt{ k^2 + \chi(0) }}
         -
         \frac{1}{\sqrt{ k^2 + m^2 }} \,
      \right ]
   \>.
\end{align}
Where we have used Eq.~\eqref{e:mtomu}.

So, for our case, Fourier transforms of the Green functions at $t=0$
are given by:
\begin{align}
   \tilde{G}_{0\,>}(k;t,t')/i
   &=
   f_k^{\phantom\ast}(t) f_k^{\ast}(t') \, ( n_k + 1 )
   +
   f_k^{\ast}(t) f_k^{\phantom\ast}(t') \, n_k \>,
   \notag \\
   \tilde{G}_{0\,<}(k;t,t')/i
   &=
   f_k^{\ast}(t) f_k^{\phantom\ast}(t') \, ( n_k + 1 )
   +
   f_k^{\phantom\ast}(t) f_k^{\ast}(t') \, n_k \>.
\end{align}
These results, together with Eq.~\eqref{e:initialfq0}, determine the
values of $\tilde{G}_{0\,>}(k;t,t')$, and all it's derivatives, at
$t=t'=0$.

%
%
\section{Numerical results}
\label{s:numerical}

We choose initial conditions for the two-point functions as
described in the last section.  In all our simulations we set the
renormalized mass parameter $m$, as defined in
Eq.~\eqref{e:mtomu}, to unity.  We have verified numerically that
our results are independent of the cut-off parameter $\Lambda$ for
values of $\Lambda$ between $3\pi$ and $4\pi$. The numerical
procedure for solving the BVA equations is described in detail in
Refs.~\cite{r:MM02,r:MS02}.  The energy is numerically conserved
to better than five significant figures for all our simulations.
The calculations are carried out entirely in momentum space, and
the results are free of artifacts related to the finite volume of
a lattice in coordinate space. For a discussion of differences
between the continuum and the periodic lattice approach,
respectively, we refer the reader to our previous paper,
Ref.~\cite{r:MD02}.

\begin{figure}[h!]
   \includegraphics[width=3.0in]{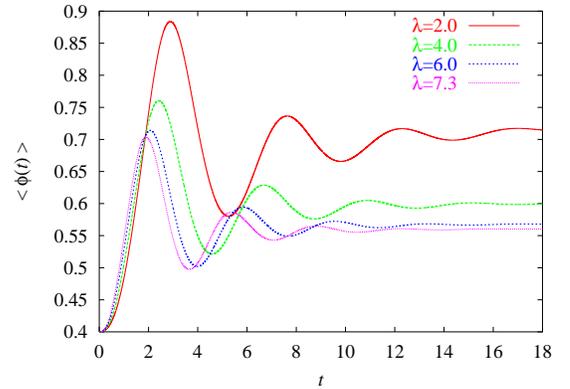}
   \caption{\label{fig:phi_lambda}Coupling constant dependence:
            We set $T_0=0.1$, and plot $\langle \phi(t) \rangle$
            for various values of $\lambda$.}
\end{figure}

\begin{figure}[h!]
   \includegraphics[width=3.0in]{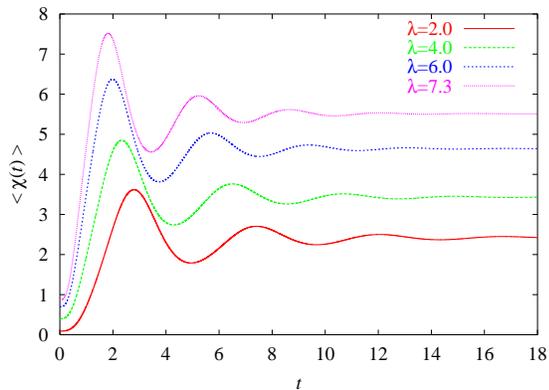}
   \caption{\label{fig:chi_lambda}Coupling constant dependence:
            We set $T_0=0.1$, and plot $\langle \chi(t) \rangle$
            for various values of $\lambda$.}
\end{figure}

We start by choosing an initial temperature $T_0 = 0.1$ so as to bring
out the quantum nature of the dynamics.  The Hartree effective
potential for this value of $T_0$ is a slowly varying function of the
coupling constant, with the minimum value of $\phi$ at large $\lambda$
being between $0.6$ and $1.0$ (see Figs.~\ref{fig:vpot_lambda} and
\ref{fig:vpot_temp}).  Thus we want to choose small initial values of
$\phi(0)=0.4$ and take $\pi(0) = \dot \phi(0)=0$ so that we start
below the height of the barrier.  Then we should see $\phi$ moving to
the opposite side of the well and then settling down at the potential
minimum position.  We show the results of this calculation in
Figs.~\ref{fig:phi_lambda} and~\ref{fig:chi_lambda} for several values
of the coupling constant.  We notice that the position of the BVA
minimum is located between $0.55$ and $0.60$ instead of the Hartree
average value of $0.8$.  We also see that the final value of $\chi$
does seem to be linear in $\lambda$ as in the Hartree result.
However, the ratio $\chi /(\lambda v^2)$ is one for the Hartree
approximation, but is about $2.4$, with a less than 1\% error, for the
BVA simulations.

\begin{figure}[h!]
   \includegraphics[width=3.0in]{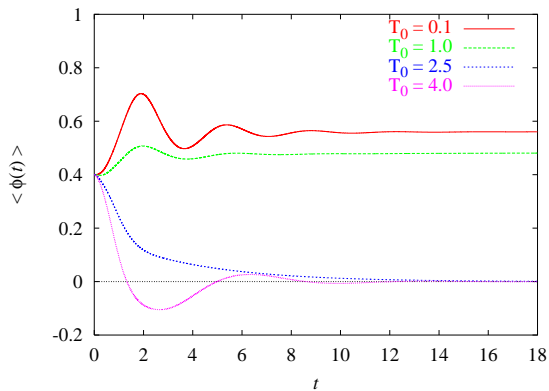}
   \caption{\label{fig:phi_temp}Temperature dependence:
            We set $\lambda=7.3$, and plot $\langle \phi(t) \rangle$
            for various values of $T_0$.}
\end{figure}

\begin{figure}[h!]
   \includegraphics[width=3.0in]{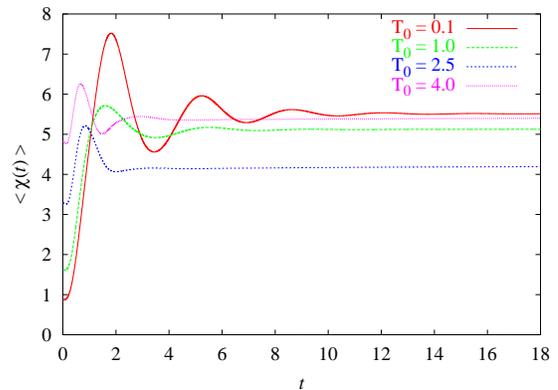}
   \caption{\label{fig:chi_temp}Temperature dependence:
            We set $\lambda=7.3$, and plot $\langle \chi(t) \rangle$
            for various values of $T_0$.}
\end{figure}

Next we set $\lambda=7.3$, which is the phenomenological choice for
the the linear sigma model, with $\phi(0)=0.4$ and $\pi(0) = \dot
\phi(0)=0$, and study the dependence of $\langle \phi(t) \rangle$ and
$\langle \chi(t) \rangle$ as a function of the initial temperature
$T_0$.  The results are shown in Figs.~\ref{fig:phi_temp} and
\ref{fig:chi_temp}.  We note immediately from Fig.~\ref{fig:phi_temp}
that for small values of $T_0$, $\langle \phi(t) \rangle$ equilibrates
to non-zero values, but for large values of $T_0$, $\langle \phi(t)
\rangle$ equilibrates to zero, as expected from the Hartree effective
potential.  Fig.~\ref{fig:chi_temp} shows that $\chi(t)$
equilibrates to different values which depend on $T_0$.

\begin{figure}[h!]
   \includegraphics[width=3.0in]{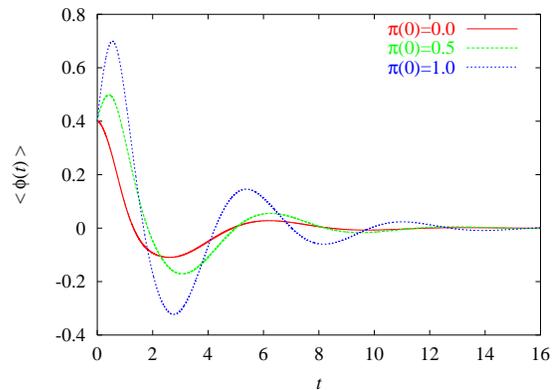}
   \caption{\label{fig:phi_speed}We are increasing the kinetic energy:
            We set $\lambda=7.3$, $T_0=0.1$, and plot
            $\langle \phi(t) \rangle$ .}
\end{figure}

Since we have chosen the case $\pi(0) = \dot \phi(0) = 0$, with
$\phi(0)$ just under the barrier height, the transition to the
equilibration point is very slow, i.e. equilibration is reached
without too many exciting features. It is interesting to give the
system a little initial kinetic energy so that structure is
introduced in the dynamics but the equilibration value of $\langle
\phi(t) \rangle$ remains the same.  We illustrate this in
Fig.~\ref{fig:phi_speed}.

\begin{figure}[h!]
   \includegraphics[width=3.0in]{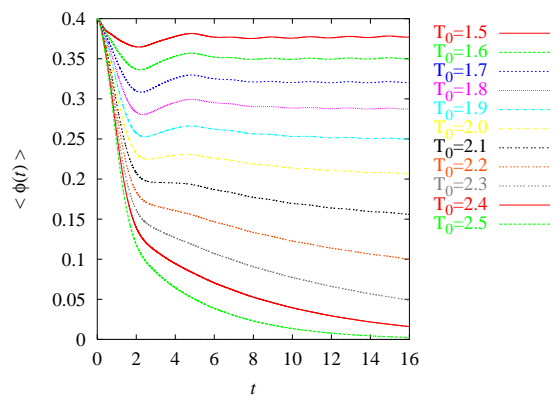}
   \caption{\label{fig:phi_temp_15_25}Detailed temperature dependence:
            We set $\lambda=7.3$, and plot $\langle \phi(t) \rangle$ .}
\end{figure}

\begin{figure}[h!]
   \includegraphics[width=3.0in]{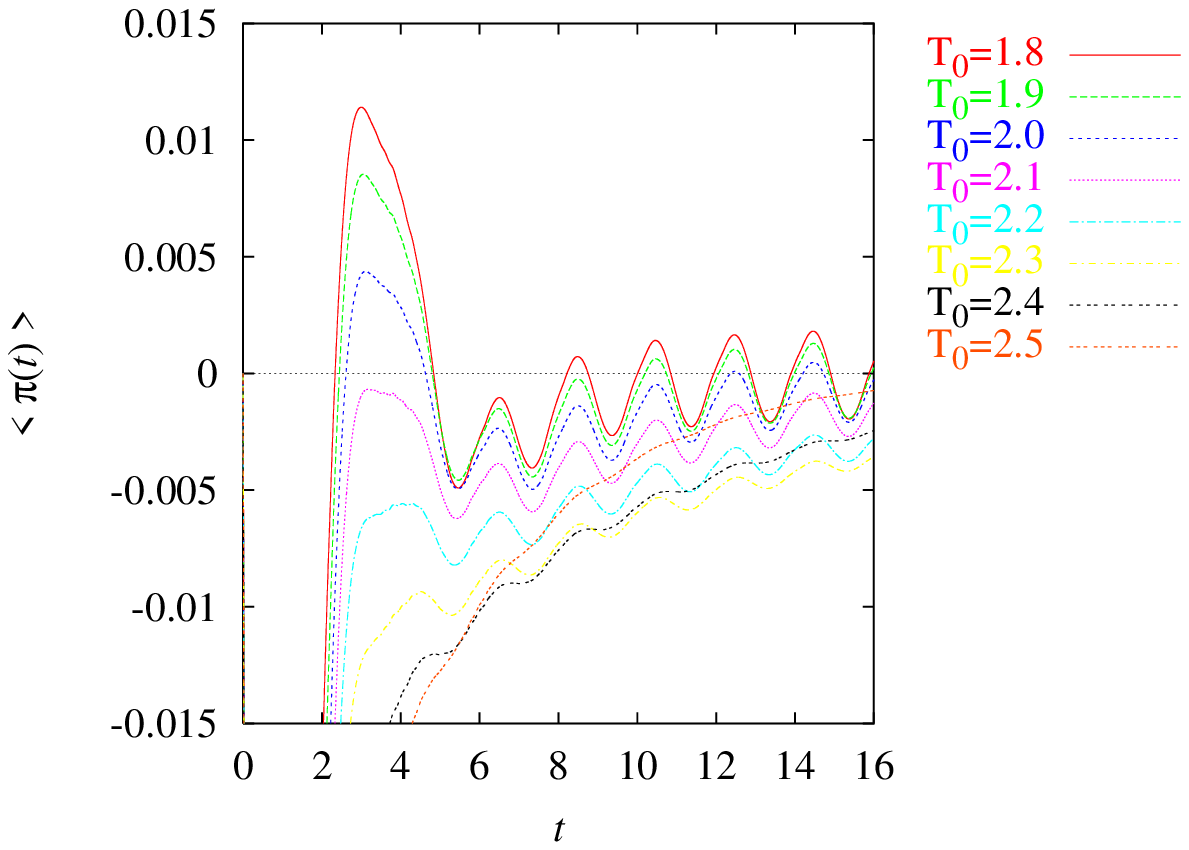}
   \caption{\label{fig:phi_dot_detail}Detailed temperature dependence:
            We set $\lambda=7.3$, and plot $\langle \dot \phi(t) \rangle$ .}
\end{figure}

The plots of the BVA order parameter in Fig.~\ref{fig:phi_temp}
indicate that for very low values of $T_0$, the order parameter
$\langle \phi(t) \rangle$ approaches a non-zero constant.  For very
large values of $T_0$, the order parameter goes to zero, as expected.
Somewhere between $T_0 = 1.0$ and $T_0 = 2.5$, there seems to be a
phase transition.  In order to study this in more detail, we have
carried out BVA simulations for temperatures between $T_0=1.5$ and
$T_0=2.5$ at $0.1$ intervals.  The results for the order parameter and
its first derivative are shown in Figs.~\ref{fig:phi_temp_15_25}
and~\ref{fig:phi_dot_detail}.

The equilibration points for the order parameter in
Fig.~\ref{fig:phi_temp_15_25} are at almost equally spaced intervals
for $T_0 < 2.4$, and there is no evidence of a jump in the value of
the order parameter from a finite value to zero, as would be required
by a first order phase transition.  The plot of $\langle \dot \phi(t)
\rangle$, shown in Fig.~\ref{fig:phi_dot_detail}, reveals the complex
nature of the dynamics: $\langle \dot \phi(t) \rangle$ oscillates and
gradually approaches zero, when equilibration occurs.  Oscillations
around a zero value of $\langle \dot \phi(t) \rangle$ suggest the
trapping in a potential well, while the wiggles at negative values of
$\langle \dot \phi(t) \rangle$ may be due to changes in the position
of the local minimum of the effective potential as a function of
temperature and time.  Since the chosen initial conditions do not
represent an equilibrium state for the interacting system, the
dynamics toward a final state of equilibrium is accompanied by changes
in the effective temperature.  In other words, the effective potential
is not such a good guide to the dynamics.  In analyzing these figures,
it becomes apparent that a) there is no abrupt change that would
support the first-order labeling of the phase transition, and b) there
is a qualitative change happening near $T_0=2.4$, where the $\langle
\dot \phi(t) \rangle$ plot begins crossing the lower temperature
curves, and the local oscillations in $\langle \dot \phi(t) \rangle$
completely disappear.  Therefore, we conclude that the phase
transition, as calculated using the BVA, is probably not first order.

\begin{figure}[h!]
   \includegraphics[width=3.0in]{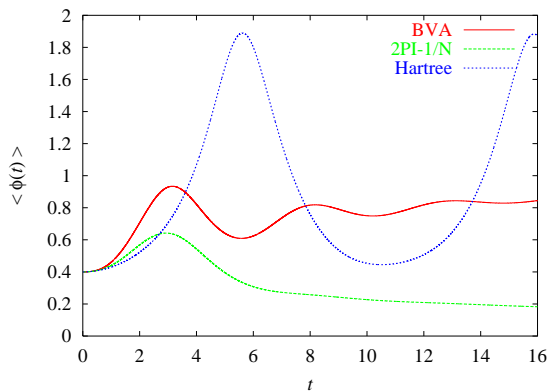}
   \caption{\label{fig:phi_comp_10}Comparison of various methods:
            We set $\lambda=1.0$ and $T_0=0.1$,
            and plot $\langle \phi(t) \rangle$.}
\end{figure}

\begin{figure}[h!]
   \includegraphics[width=3.0in]{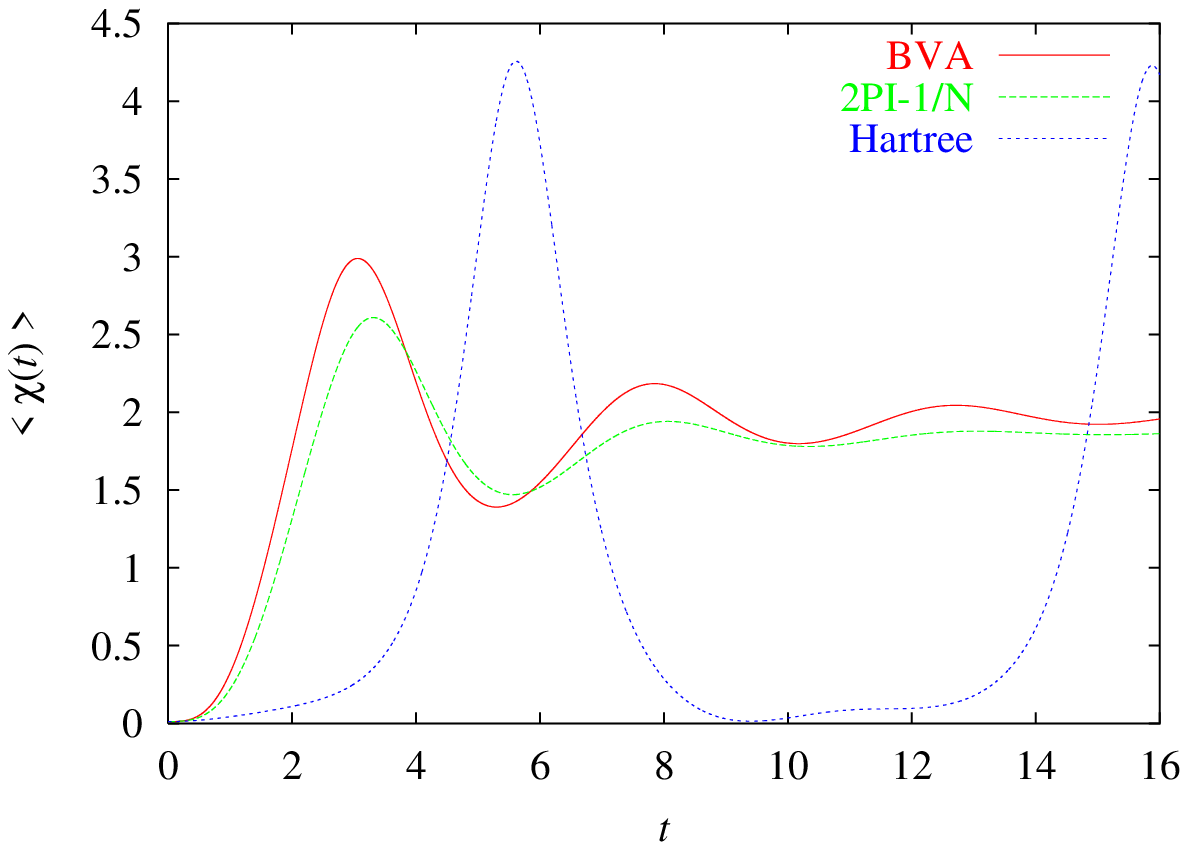}
   \caption{\label{fig:chi_comp_10}Comparison of various methods:
            We set $\lambda=1.0$ and $T_0=0.1$,
            and plot $\langle \chi(t) \rangle$.}
\end{figure}

\begin{figure}[h!]
   \includegraphics[width=3.0in]{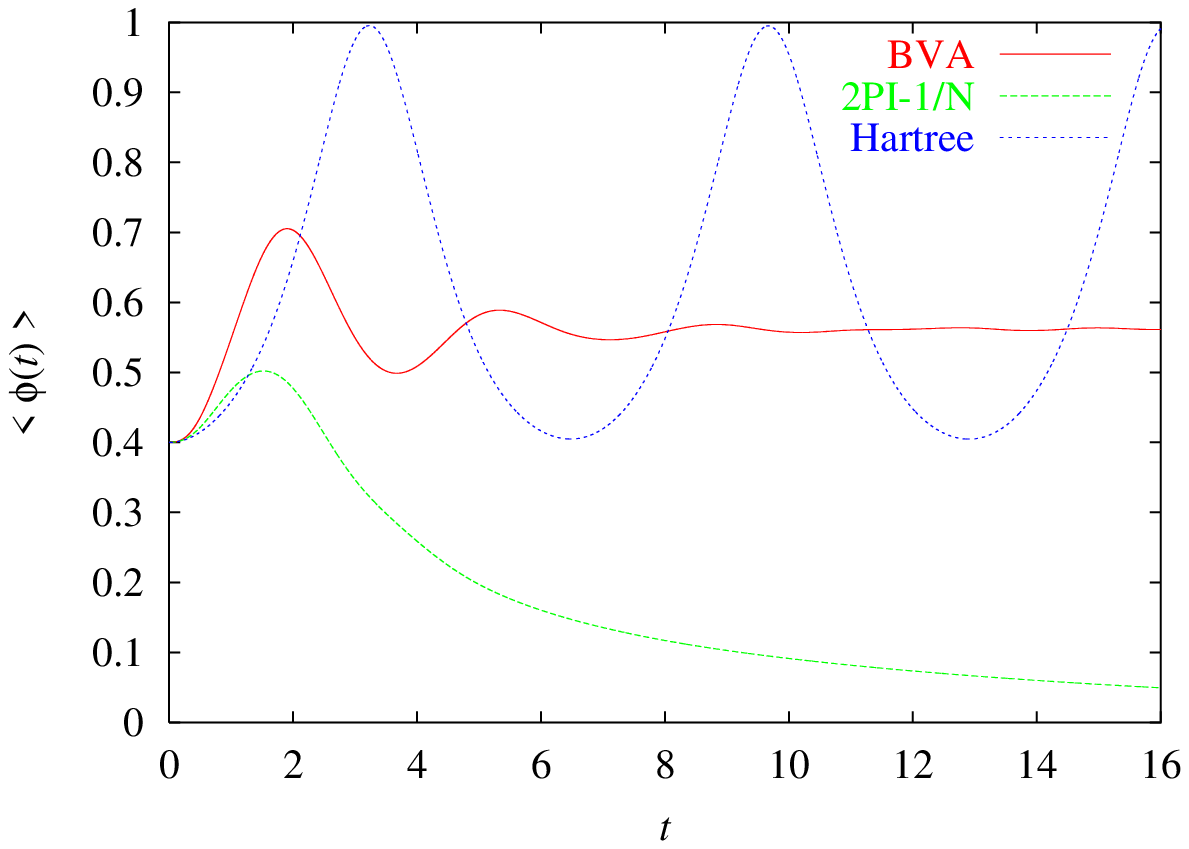}
   \caption{\label{fig:phi_comp_73}Comparison of various methods:
            We set $\lambda=7.3$ and $T_0=0.1$,
            and plot $\langle \phi(t) \rangle$.}
\end{figure}

\begin{figure}[h!]
   \includegraphics[width=3.0in]{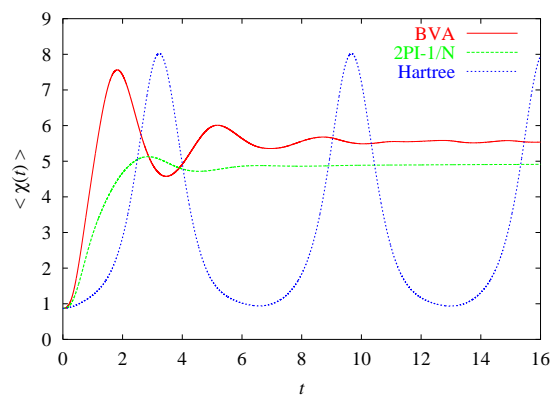}
   \caption{\label{fig:chi_comp_73}Comparison of various methods:
            We set $\lambda=7.3$ and $T_0=0.1$,
            and plot $\langle \chi(t) \rangle$.}
\end{figure}

It is known that this model in two-dimensions has no phase
transition at finite temperature~\cite{r:BrezinZinnJustin}.  At
zero temperature, there is a second order phase
transition~\cite{r:Chang76,r:Drell77,r:Banks78}. Explicit Monte
Carlo lattice calculations~\cite{r:amaral85,r:amaral86} have shown
that indeed $\lambda \phi^4$ theory in two-dimensions and at zero
temperature is non-trivial  at least when the continuum limit is
reached from the broken symmetry phase, and that the symmetry is
fully restored at high temperature. It is known however that
approximate lattice calculations (such as the variational-cumulant
expansion method~\cite{r:LiChen95,r:LiChen96}), which are designed
to study scalar $\phi^4$ theory in 3+1 dimensions on the lattice,
may erroneously indicate the presence of a second-order phase
transition at finite temperature in 1+1 dimensions. This is
probably due to the fact that the expansion is only carried out to
third order.
Much in the same way, the Hartree approximation exhibits a first
order phase transition with $T_{\text{cr}} \approx 0.878$. When
going beyond the Hartree approximation (using the BVA) we still
find a phase transition, but the BVA relaxes the
order 
of the phase transition.
At very low temperature, we do (apparently) find a non-zero value
of the order parameter as $t \rightarrow \infty$, which is the
exact result at zero temperature.  So we conclude that the BVA at
low temperature is seeing effects that might occur in higher
dimensions, even though it is not technically correct in 1+1
dimensions.

We next compare three different approximation methods: Hartree, the
2PI-1/N expansion, and BVA.  Here we choose a very low initial
temperature of $T_0=0.1$ in order to emphasize quantum effects in the
dynamics.  Again, we start with $\langle \phi(0) \rangle = 0.4$ and
$\langle \dot{\phi}(0) \rangle = 0$.  We show results for $\lambda =
1$ in Figs.~\ref{fig:phi_comp_10} and~\ref{fig:chi_comp_10} and
$\lambda = 7.3$ in Figs.~\ref{fig:phi_comp_73}
and~\ref{fig:chi_comp_73}.  As expected, we find that the Hartree
approximation leads to oscillation about the Hartree minimum without
equilibration.  The BVA results track the Hartree curve, except with
damping, and go to a non-zero value as $t \rightarrow \infty$, which
is the exact result at zero temperature.  The 2PI-1/N expansion goes
to a zero value of the order parameter, in agreement with the exact
result of no phase transition for finite temperature.  Results for
$\langle \chi(t)\rangle$ are similar for both approximations, in
agreement with expectations based on our experience with the classical
limit of these approximations~\cite{r:CDM02}.  The disagreement
between 2PI-1/N and BVA is more pronounced at larger values of
$\lambda$, as shown in Figs.~\ref{fig:phi_comp_73}
and~\ref{fig:chi_comp_73}.

%
%
\section{Thermalization}
\label{s:thermalization}

The BVA leads to thermalization of the system.  In order to have a
measure of the semi-static thermodynamic properties of the system,
it is reasonable to fit our time-dependent Green functions to
those appropriate to a free field with frequency $\omega_k(t)$ and
Bose-Einstein distribution function $n_k(t)$, see also
\cite{r:AB01}. That is, we equate the Fourier transform of the BVA
Green functions $\tilde{G}_{k}(t,t)$ and $\partial^2
\tilde{G}_k(t,t') /
\partial t \, \partial t'$ to the corresponding free-field cases:
\begin{gather}
   \tilde{G}_{k}(t,t)
   =
   [  2 n_k(t) + 1 ] \,
   \frac{ 1 }{ 2 \omega_k(t) }  \>,
   \\
   \left .
      \frac{ \partial^2 \tilde{G}_k(t,t') }{ \partial t \, \partial t' }
   \right |_{t=t'}
   =
   [ 2 n_k(t) + 1 ] \,
   \frac{ \omega_k(t) }{2}  \>,
   \label{e:Gtofree}
\end{gather}
where we set
\begin{gather}
   n_k(t)
   =
   \frac{ A(t) }
   { e^{ \omega_k(t) / T_{\text{eff}}(t) } - 1 }
   \>,
   \notag \\
   \omega_k(t)
   =
   \sqrt{ k^2 + m^2_{\text{eff}}(t) } \>,
   \label{e:neffomegaeff}
\end{gather}
with $m^2_{\text{eff}}(t)$ the effective mass and $T_{\text{eff}}(t)$,
the effective temperature at time $t$.  The factor $A(t)$ comes from
a wave function renormalization at each time $t$.  So we can determine
$n_k(t)$ and $\omega_k(t)$ from the relations:
\begin{gather}
   \omega_k(t) = \left [
        \left . \frac{ \partial^2 G(t,t';k) }{ \partial t \, \partial t' }
        \right |_{t=t'}
        \ \Big / \
        G(t,t;k) \right ]^{\frac{1}{2}}
   \>,
   \label{eq:omegak}
   \\
   n_k(t) = \frac{1}{2} +
   \left [
        \left . \frac{ \partial^2 G(t,t';k) }{ \partial t \, \partial t' }
        \right |_{t=t'}
        \
        G(t,t;k) \right ]^{\frac{1}{2}}
   \>.
   \label{eq:nk}
\end{gather}
>From \eqref {eq:omegak}, we notice that $\omega_k$ is a ratio of
Green functions, and thus any (finite) wave function
renormalization will cancel.  However from \eqref{eq:nk}, $n_k$ is
directly proportional to the Green function, which will have a
finite wave function renormalization when restricted to the single
particle contribution.

\begin{figure}[h!]
   \includegraphics[width=3.0in]{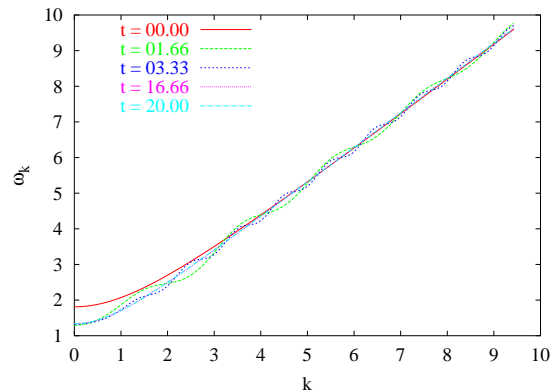}
   \caption{\label{fig:omega_k_73_25} For $\lambda=7.3$ and $T_0=2.5$
            we plot the time evolution of $\omega_k$.}
\end{figure}

\begin{figure}[h!]
   \includegraphics[width=3.0in]{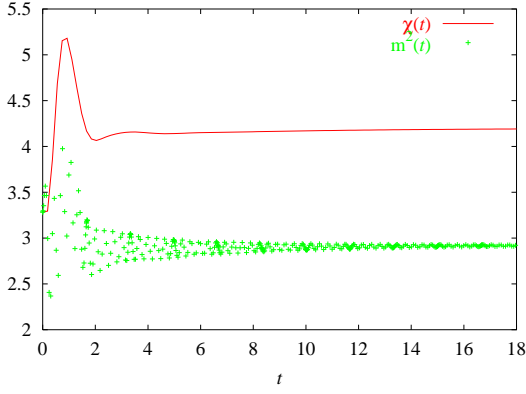}
   \caption{\label{fig:m2_73_25} For $\lambda=7.3$ and $T_0=2.5$
            we compare $m^2(t)$ and $\chi(t)$.}
\end{figure}

First let us concentrate on $\omega_k(t)$.  We show in
Fig.~\ref{fig:omega_k_73_25} a plot of $\omega_k(t)$ as a function of
$k$ for various values of $t$, as calculated from
Eq.~\eqref{eq:omegak}, for $\lambda=7.3$ and $T_0=2.5$, which is close
to the critical temperature.  We fit the $\omega_k(t)$ data to
determine the values of $m^2_{\text{eff}}(t)$.  These results are
compared to the value of $\chi(t)$ in Fig.~\ref{fig:m2_73_25}.  We see
here that self-energy corrections to the effective mass reduce the
effective mass to about 25-40\% of $\chi(t)$.  Further calculations
for different initial temperatures allow us to conclude that the
correction slowly decreases with increasing initial temperature.

\begin{figure}[h!]
   \includegraphics[width=3.0in]{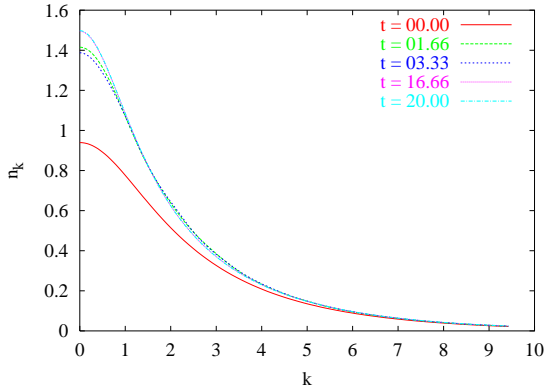}
   \caption{\label{fig:n_k_73_25} For $\lambda=7.3$ and $T_0=2.5$
            we plot the time evolution of $n_k$, before renormalization.}
\end{figure}

Now that we have the effective $\omega_k$ we can find out if the
particle number density $n_k(t)$ has the simple Bose-Einstein form
given in Eq.~\eqref{e:neffomegaeff}.  We use a nonlinear fitting
procedure to obtain the parameters $T_{\text{eff}}(t)$ and $A(t)$
from the data generated by Eq.~\eqref{eq:nk}.  The results are
shown in Fig.~\ref{fig:n_k_73_25} for $\lambda=7.3$ and $T_0=2.5$.
In Fig.~\ref{fig:Tk_ren_73_25} we show the \emph{renormalized}
densities, given by $n_{\text{ren}\,k}(t) = n_k(t) / A(t)$. Notice
that $n_k(t)$ is very similar to a Bose-Einstein distribution
which starts out at a temperature of $T_{\text{eff}} = T_0$,
increases in amplitude due to a temperature increase, then falls
back to a lower temperature, but larger mass.

\begin{figure}[h!]
   \includegraphics[width=3.0in]{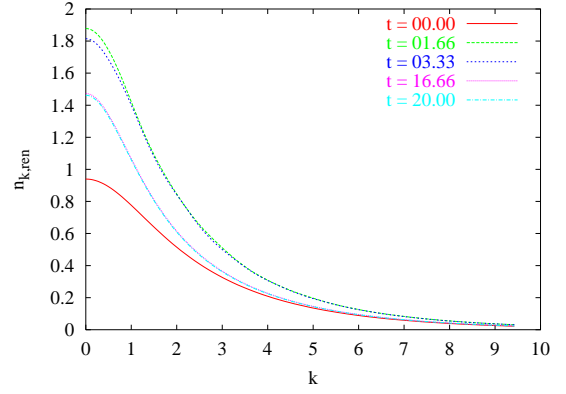}
   \caption{\label{fig:nk_ren_73_25} For $\lambda=7.3$ and $T_0=2.5$
            we plot the time evolution of $n_{k,\text{ren}}$, after
             renormalization.}
\end{figure}

\begin{figure}[h!]
   \includegraphics[width=3.0in]{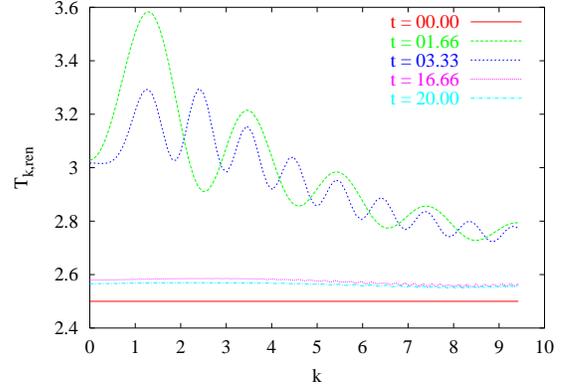}
   \caption{\label{fig:Tk_ren_73_25} For $\lambda=7.3$ and $T_0=2.5$
            we plot the time evolution of the temperature
            $T_{k,\text{ren}}(t)$.}
\end{figure}

In Fig.~\ref{fig:Tk_ren_73_25} we show the effective temperature
profile, $T_{k,\text{ren}}(t)$ inferred from $n_{k,\text{ren}}(t)$
\begin{equation}
   T_{\text{ren}\,k}(t)
   =
   \omega_k(t) \,
   / \,
   \ln
   \bigl [ \,
       ( n_{\text{ren}\,k}(t) + 1 ) /
       ( n_{\text{ren}\,k}(t) ) \,
   \bigr ]
   \>.
\end{equation}
We notice that at short times, $T_{\text{ren}\,k}(t)$ has to readjust to
the effects of the non-Gaussian corrections, but then it settles down
to a new temperature which is relatively independent of the momentum
$k$. We also notice that we do have quite a bit of particle
production: after an initial spike, the particle density number
relaxes to an equilibrium value.

\begin{figure}[h!]
   \includegraphics[width=3.0in]{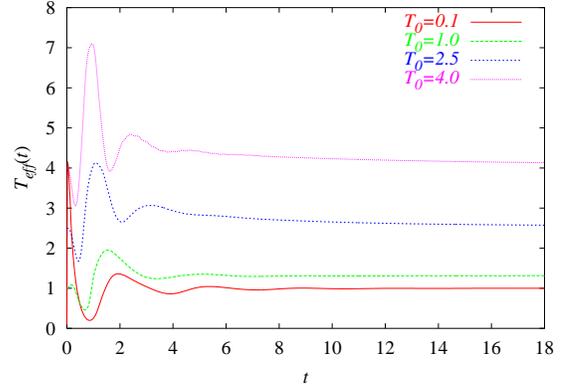}
   \caption{\label{fig:Teff_73} For $\lambda=7.3$ we plot the time
            evolution of $T_{\text{eff}}(t)$ for various initial
            temperatures $T_0$.}
\end{figure}

\begin{figure}[h!]
   \includegraphics[width=3.0in]{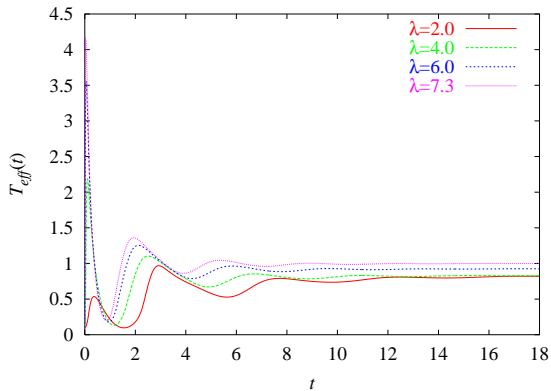}
   \caption{\label{fig:Teff_01} For $T_0=0.1$ we plot the time
            evolution of $T_{\text{eff}}(t)$ for various values of the
            coupling constant $\lambda$.}
\end{figure}

In Fig.~\ref{fig:Teff_73} we show the dependence of
$T_{\text{eff}}$ on the initial temperature $T_0$, for fixed
$\lambda = 7.3$, and in Fig.~\ref{fig:Teff_01}, the dependence of
$T_{\text{eff}}$ on the coupling constant $\lambda$ for fixed
$T_0=0.1$.
%
%
\section{Conclusions}
\label{s:conclusions}

In conclusion, we have shown that the BVA indeed leads to
equilibration of the one- and two-point functions and thus remedies
this deficiency of the Hartree approximation.  The nature of the phase
transition changes from first to second order when going from Hartree
to BVA which is the correct behavior (as a function of coupling
constant) at zero temperature but is incorrect at higher
temperature. However, such a phase transition at finite temperature is
expected in the 3+1 dimensional case that we are interested in
modeling in the future.  The 2PI-1/N approximation does not see this
phase transition (which is not present in the exact theory) and does
not track the average of the Hartree result.  In higher dimensions,
one expects the Hartree result for the order parameter to be correct
on the average but not have the property of equilibrating.  So
deciding which approximation is more physical will have to wait for a
3+1 dimensional simulation to see if both approximations show a phase
transition.

In the classical regime, where we could do an exact Monte Carlo
calculation, we found that the BVA works better than the 2PI-1/N
expansion in capturing the dynamics of $\langle \phi(t) \rangle$,
but that the difference was not very great.  In the quantum
domain, however, these two approximations diverge at low
temperature with the strength of the coupling constant, and the
2PI-1/N approximation no longer tracks the average of the results
of the Hartree approximation. Since we do not have exact
calculations in the quantum regime, we cannot make any strong
conclusions about this divergence.

It is interesting to note that technical issues related to the
nature of the phase transitions in 1+1 dimensions are not
uncommon. There is a previous example of just this very type of
reasoning being successful.  In trying to understand the QCD
chiral phase transition at finite chemical potential and
temperature, it was important to have a model where the phase
transition mimicked what is known in 3+1 dimensions.  Such a model
was a four-fermi model (the Gross-Neveu model) in 1+1 dimensions
in leading order in large-N, which had a similar phase structure
to two-flavor QCD.  This model was then used as a testing ground
for studying the effect of first and second order phase
transitions, with a critical point, and showed some qualitative
differences between the two types of transitions (see
Ref.~\cite{r:CCMS01}).  Once again, the exact 1+1 dimensional
model does not have a phase transition.  But this warmup problem
then allowed us to go to 2+1 dimensions\cite{r:CS02}, where it is
known that the leading order large-N approximation and the exact
theory have similar phase diagrams, as verified by lattice
simulations in 2+1 dimensions~\cite{r:HGS01}.  We submit that this
is the way these 1+1 dimensional results should be understood: as
a model having certain properties when treated in this
approximation and as a testing ground for codes which will then be
generalized to higher dimensions where it is expected the
approximation will correspond to the known behavior. The main
purpose of the quantum simulations we present here is to get
experience in getting codes working in lower dimensions that
qualitatively do what we expect to see in 3+1 dimensions.

As a result of the simulations presented here, we are confident that
we can now study the chiral phase transition in 3+1 dimensions in the
linear sigma model and describe the competition between the expansion
of the plasma and the equilibration tendencies.  This will allow us to
see whether some of the phenomena present in Hartree (and/or large-N)
approximation, such as production of disoriented chiral condensates
and distortion of pion and dilepton spectra, are still present in
spite of the forces that lead to thermalization.

%
%

\begin{acknowledgments}

Numerical calculations are made possible by grants of time on the
parallel computers of the Mathematics and Computer Science
Division, Argonne National Laboratory. The work of BM was
supported by the U.S. Department of Energy, Nuclear Physics
Division, under contract No. W-31-109-ENG-38.  JFD and BM would
like to thank Los Alamos National Laboratory and the Santa Fe
Institute for hospitality.

\end{acknowledgments}
%
%
%
\bibliography{johns}
%
%
\end{document}